\documentclass[aps,prl,twocolumn,floatfix,superscriptaddress]{revtex4-1}
\usepackage{graphicx,bm}
\usepackage{times}
\usepackage{amsmath,amssymb}
\usepackage{xcolor}

\newcommand{\beginsupplement}{%
	\setcounter{table}{0}
	\renewcommand{\thetable}{S\arabic{table}}%
	\setcounter{figure}{0}
	\renewcommand{\thefigure}{S\arabic{figure}}%
}

\begin{document}

\title{Tunable current circulation in triangular quantum-dot metastructures}

\author{Chen-Yen Lai}
\affiliation{Theoretical Division, Los Alamos National Laboratory, Los Alamos, New Mexico 87545, USA}
\affiliation{Center for Integrated Nanotechnologies, Los Alamos National Laboratory, Los Alamos, New Mexico 87545, USA}
\author{Massimiliano Di Ventra}
\affiliation{Department of Physics, University of California, San Diego, La Jolla, CA 92093, USA.}
\author{Michael Scheibner}
\affiliation{School of Natural Sciences, University of California, Merced, Merced, CA 95343, USA.}
\author{Chih-Chun Chien}
\email{cchien5@ucmerced.edu}
\affiliation{School of Natural Sciences, University of California, Merced, Merced, CA 95343, USA.}

\date{\today}

\begin{abstract}
Advances in fabrication and control of quantum dots allow the realization of metastructures that may exhibit novel electrical transport phenomena.
Here, we investigate the electrical current passing through one such metastructure, a system composed of quantum dots placed at the vertices of a triangle. The wave natural of quantum particles leads to internal current circulation within the metastructure in the absence of any external magnetic field.
We uncover the relation between its steady-state total current and the internal circulation.
By calculating the electronic correlations in quantum transport exactly, we present phase diagrams showing where different types of current circulation can be found as a function of the correlation strength and the coupling between the quantum dots.
Finally, we show that the regimes of current circulation can be further enhanced or reduced depending on the local spatial distribution of the interactions, suggesting a single-particle scattering mechanism is at play even in the strongly-correlated regime.
We suggest experimental realizations of actual quantum-dot metastructures where our predictions can be directly tested.
\end{abstract}

\pacs{
}

\maketitle

Quantum transport is an important area of research with a wide range of phenomena and applications, especially in fields like condensed-matter~\cite{datta2005quantum,DiVentra:2008ks} and cold-atom systems~\cite{Lai:2017kq,Amico:2015tk,Olsen:2015hd,Chien:2015kc,Lai:2016kh}.
Of particular present interest are those phenomena that emerge in the presence of nontrivial geometry or topology.
The prototypical example is the Aharonov-Bohm (AB) effect~\cite{AB59} which arises when a finite vector potential is encircled by a conducting ring and endows the electron wave-function with an additional geometrical phase.
However, several other interesting transport phenomena emerge from geometry and topology, such as quantized conduction via edge states of topological insulators that have been an important probe for nontrivial topology in the band structure~\cite{Kane_TIRev,Zhang-TIRev,ShenTI,Asboth2016}, or flat bands of geometrically-induced localized states that interfere with mobile particles and influence their transport~\cite{Metcalf:2015wj,Lai:2016hr}, to name just a few.

Nanoscale structures, such as quantum dots (QDs), offer additional opportunities to engineer {\it metastructures} that, if appropriately constructed, may reveal quantum transport phenomena otherwise difficult to probe with other means.
Here, we investigate quantum currents through a topologically nontrivial metastructure consisting of QDs placed at the vertices of a triangle with additional elements for tuning the tunneling and interactions.
We call it a ``triangular quantum-dot metastructure'' (TQDM).
These metastructures resemble the triangular triple quantum dot, which has been already fabricated and employed in studying other physical properties~\cite{Seo13,Mitchell10,Kim15,Noiri17,Zhang18} (see below).

To probe the internal electrical dynamics of the TQDM, we connect two of the three QDs to two external reservoirs, as illustrated in Fig.~\ref{fig:illu}(a).
Such a system forces the currents to flow through a non simply-connected region, generating current circulation within the TQDM without the need of a vector or scalar potential.
The internal circulation is possible because of the non-trivial topology of the TQDM and the wave nature of quantum particles. The quantum current on one path may overshoot the total current, so the other path flows reversely to compensate. 
In addition, one can detect the emergence of the internal TQDM current circulation by varying a single link between two of the three QDs, giving rise to a non-monotonic behavior of the total current.

By introducing correlations one can tune the circulation further by switching from clockwise (CW) to counterclockwise (CCW) to no-circulation or unidirectional (UD) flow.
Computing electronic correlations {\it exactly} in the Hubbard model, we provide the corresponding phase diagrams of current circulation as a function of correlation strength and inter-dot coupling. We also study the effect of inhomogeneous correlations, that reveal a single-particle scattering mechanism is at play even in the strongly-correlated regime.
We report here the results obtained using an open-system, quantum master equation approach.
In the Supplemental Information, we report those obtained by a microcanonical (closed-system) formalism~\cite{DiVentra:2004bx}, showing that the two approaches lead to the same conclusions.
The agreement establishes the model-independence of the internal circulation of current in a multi-connected geometry.

\begin{figure}[t]
  \begin{center}
    \includegraphics[width=1\columnwidth]{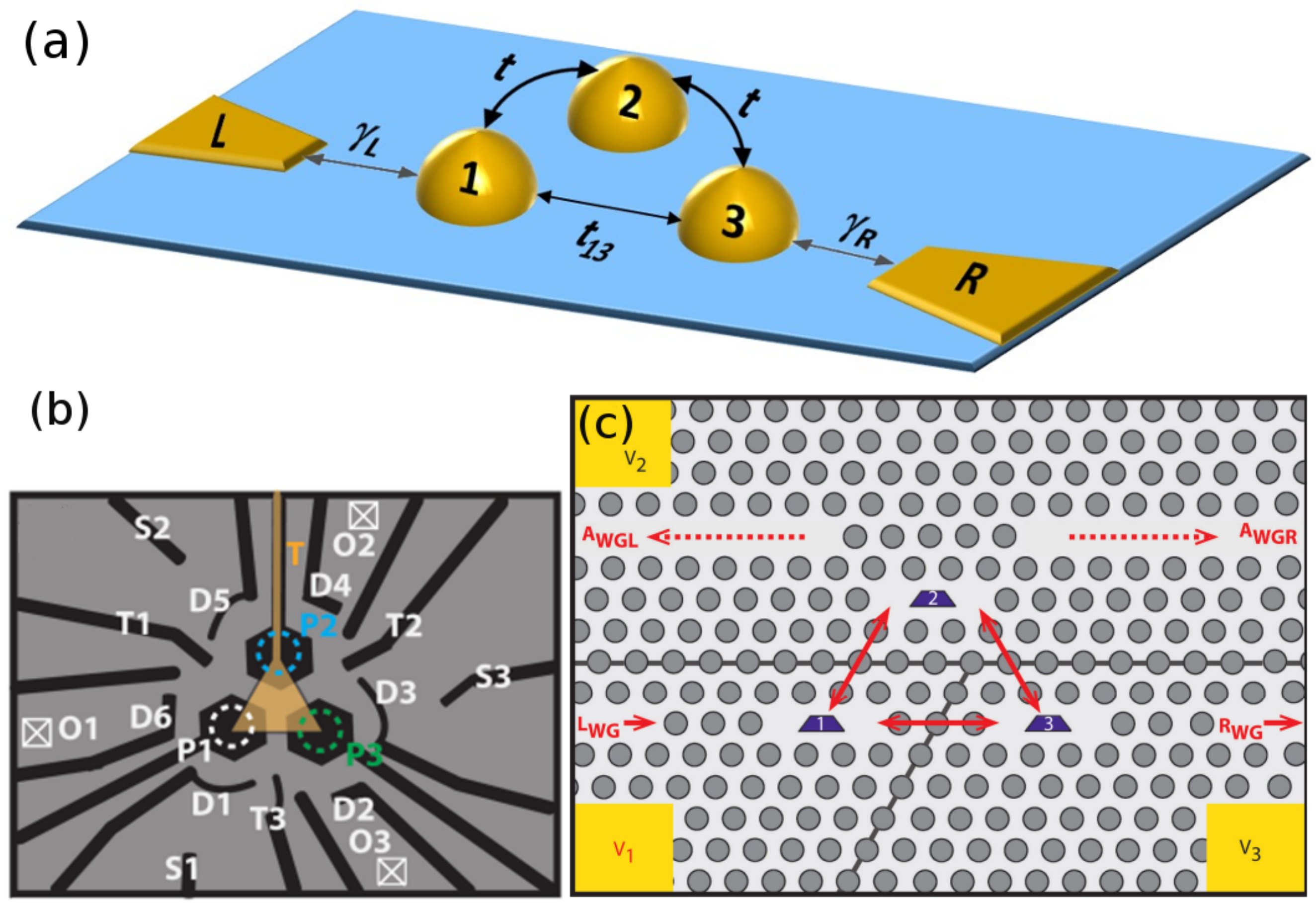}
    \caption{
			(a) Schematic rendering of a triangular quantum-dot metastructure (TQDM) connected to two reservoirs for studying internal current circulation.
			Here the two reservoirs labeled "L" and "R" are coupled to the TQDM via the coupling $\gamma_{L,R}$, respectively. The tunneling coefficients $t$ and $t_{13}$ are assumed to be tunable.
			(b) Schematic of a possible experimental structure of a TQDM formed electrostatically in a 2DEG.
			The design is adapted from~\cite{Noiri17}.
			Here, gates T1-T3 control the inter-dot tunneling between dots P1-P3, the (orange) top gate T establishes a depletion region in the center between the dots, gates D1-D6 define the outer boundaries of the three dots, and the additional gates S1-S3 and other contacts can be used for charge sensing.
			O1-O3 represent Ohmic contacts which serve as leads to/from the QDs.
			(c) Schematic of possible experimental opto-electronic realization of the TQDM system.
			The dots are embedded in three L3 cavities in a photonic crystal membrane, formed by a pin-type diode.
			Photons are injected from the left waveguide ($L_{WG}$) and extracted from the right waveguide ($R_{WG}$).
			Additional wave guides ($A_{WGL}$, $A_{WGR}$) may be used to measure the directionality of the photon flux.
			The three dot-cavity systems can be electrically separated from each other by etching through the top p-doped layer of the pin-diode structure.
			This allows the application of different electric fields via gate voltages $V_1$-$V_3$, thereby allowing individual tuning of the QD transitions and coupling strengths.
    }
    \label{fig:illu}
  \end{center}
\end{figure}

\textit{Experimental realization -} Before embarking on the theoretical aspects of TQDMs, let us first point out how they can be engineered with the appropriate features to observe the phenomena we predict.
First, we note that triangular triple quantum dots have been fabricated to study various phenomena such as charge frustration~\cite{Seo13} and tunable transport~\cite{Noiri17}.
There are also proposals of using the TQDM to study quantum phase transitions~\cite{Mitchell10,Kim15}, and a thermal transistor where a triangular triple quantum dot is coupled to three reservoirs has also been proposed~\cite{Zhang18}.

TQDMs can be experimentally realized in several ways.
The most obvious one relating to electronic transport utilizes electrostatically defined QDs.
Here the dots, the barriers between the dots, and the source and drain are controlled by electronic gates which modify the potential landscape of a 2-dimensional electron gas (2DEG).
The gates may be created by electron beam lithography~\cite{Noiri17} or local anodic oxidation~\cite{Ishii95,Held99,Keyser00,Rogge08} on top of an epitaxially grown 2DEG semiconductor heterostructure.
Figure~\ref{fig:illu}(b) illustrates this approach.

As another experimental realization, we propose a photonic circuit architecture, operated in the photon blockade regime to achieve the required fermionic behavior~\cite{Bimbaum05}.
An example design is shown in Fig.~\ref{fig:illu}(c).
Here, the QDs are embedded in three L3 photonic crystal cavities, which are spatially arranged to create the triangular topology.
The source and drain are formed by two photonic crystal waveguides, which guide photons from and to the input and output couplers and couple to the excitons in the quantum dots $1$ and $3$.
Auxiliary waveguides, weakly coupled to dot number 2, may be used to measure the directionality of the ``photon'' current simulating the electronic current.
The fermion spin may be simulated by coupling polarized photons to form polaritons.
Coupling between the three cavities depends on the separation, orientation and structural details. (See the SI for more details.)

\textit{Theoretical model -} To describe the steady-state transport of fermions through the TQDM, we employ an open-system approach and solve a Markovian quantum master-equation by considering a triangular lattice whose site-$1$ and site-$3$ are connected to two particle reservoirs via couplings $\gamma_L$ and $\gamma_R$, as illustrated in Fig.~\ref{fig:illu}(a).
The left (right) reservoir acts as a particle source (drain) which pumps (removes) particles into (out of) the triangle.
Here, we make the assumptions that the coupling between the system and reservoirs is weak in the sense that the frequency scale associated with the coupling between the system and environment is small compared to the dynamical frequency scales of the system or the reservoirs.
Moreover, the Markovian approximation requires the coupling to be time-independent and the time evolution of the TQDM to be slow compared to the time necessary for the environment to ``forget'' quantum correlations~\cite{breuer2007theory,weiss2012quantum}.

Then, the dynamics can be described by the Lindblad equation ($\hbar =1$ throughout):
\begin{eqnarray}\label{eq:master}
	\frac{d\rho}{dT}&=&i\left[\rho,\mathcal{H}\right]
	+\gamma_L\left(c^\dagger_1\rho c_1-\frac{1}{2}\{c_1c^\dagger_1,\rho\}\right) \nonumber \\
	&&+\gamma_R\left(c_3\rho c^\dagger_3-\frac{1}{2}\{c^\dagger_3c_3,\rho\}\right),
\end{eqnarray}
where $\rho$ is the density matrix of the TQDM and $\{A,B\}$ denotes the anticommutator of $A$ and $B$.

Here we assume the three QDs have identical energy levels and focus on transport through a single level at the Fermi energy.
The effects of inhomogeneous energy levels will be discussed later.
By assuming large energy gaps between the energy levels, we choose as Hamiltonian, $\mathcal{H}$, of the TQDM that of a single-band triangular lattice:
\begin{equation}
	\mathcal{H}=\sum_\sigma\mathcal{H}_{\text{tri},\sigma}+\sum_{p=1}^{3}U_pn_{p\uparrow}n_{p\downarrow},\label{totalH}
\end{equation}
where
\begin{equation}\label{eq:Htri}
	\mathcal{H}_{\text{tri},\sigma}=-t(c^\dagger_{1\sigma}c_{2\sigma}+c^\dagger_{2\sigma}c_{3\sigma}+h.c.)-t_{13}(c^\dagger_{1\sigma}c_{3\sigma}+c^\dagger_{3\sigma}c_{1\sigma}).
\end{equation}
Here, $c^\dagger_{n\sigma}$ ($c_{n\sigma}$) is the fermion creation (annihilation) operator on site $n$ and $\sigma\!=\uparrow,\downarrow$ denotes the spin.
The number operator of the spin $\sigma$ fermions on site $p$ is  $n_{p\sigma}\!=\!c^\dagger_{p\sigma}c_{p\sigma}$.
The coupling between QD 1 and 3 is labeled $t_{13}$, which can be tuned independently of the other coupling, $t$, between QDs 1 and 2, and QDs 2 and 3.
The onsite interaction with the coupling constant $U_p$ models the Coulomb interaction and additional coupling to the photonic structure, which may also be tuned for each QD independently.
The time unit is $T_0\equiv \hbar/t$.

The current operator from site $p$ to site $q$ is given by
\begin{equation}
	\hat{j}_{pq}=-i\sum_{\sigma}(t_{pq}c^{\dagger}_{p\sigma} c_{q\sigma} - h.c.).
\end{equation}
Here $t_{pq}$ is the hopping coefficient from $p$ to $q$.
In this work we consider the zero-temperature limit.
The current on the link from site $p$ to site $q$ of the triangular lattice can be extracted from the off-diagonal elements of the single-particle correlation matrix $\langle C_{pq\sigma}\rangle\!=\!\langle c^{\dagger}_{p\sigma} c_{q\sigma}\rangle$ as $j_{pq}\!=\!-2\sum_{\sigma}\text{Im}\langle t_{pq}C_{pq\sigma}\rangle$. In the following we choose $\gamma_L\!=\!\gamma_R\!=\!\gamma$ and focus on the steady state where $d\rho/dT\!=\!0$ in the long-time limit ($T\!\rightarrow\!\infty$), and a steady-state current can be identified. We now analyze both the total current flowing through the triangle and the internal currents inside the triangle, by varying $t_{13}$ within the triangle, the system-reservoir coupling $\gamma$, and the strength of the interaction $U_p$. Details of the calculations can be found in the SI as well as confirmation of these results using a microcanonical approach.

\begin{figure}[t]
	\begin{center}
		\includegraphics[width=0.4\textwidth]{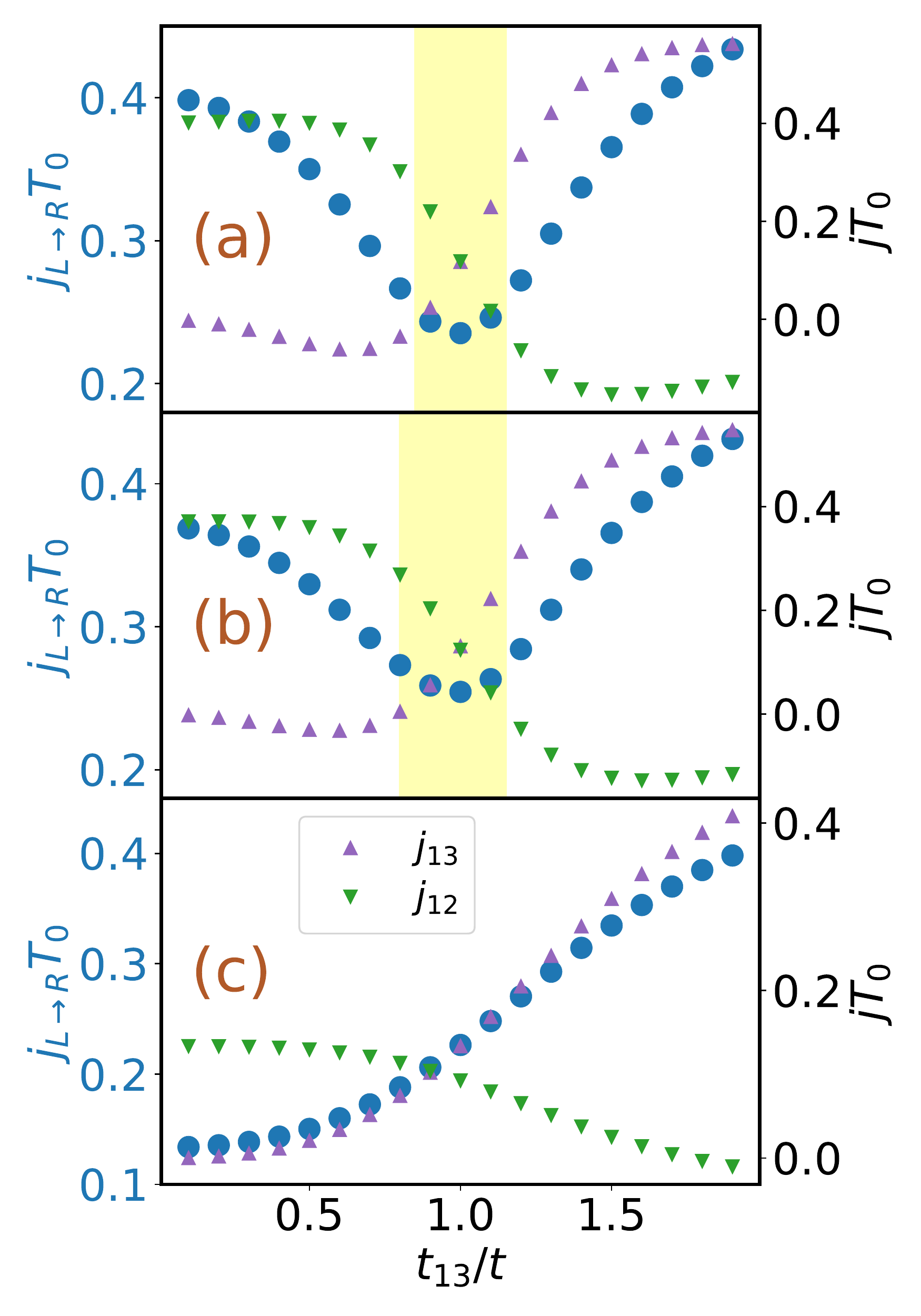}
		\caption{
			Steady-state currents versus different hopping coefficient $t_{13}$ of the triangular lattice connected to two reservoirs modeled as an open quantum system with $\gamma T_0\!=1$ and different interaction strengths (a) $U\!=\!0$, the noninteracting case, (b) $U\!=\!1t$, and (c) $U\!=\!5t$.
			Here $T_0\!\equiv\!\hbar/t$.
			The solid circles show the total current through the triangle, and the triangle symbols (upside-down triangle symbols) show the internal currents through the 1-3 (1-2) link.
			The shaded regions  in (a) and (b) emphasize the non-monotonic behavior of the total current.
			In the shaded regions all the internal currents in the triangle are uni-directional (no circulation).
		}
		\label{fig:OSIntJtotal}
	\end{center}
\end{figure}

\textit{Noninteracting fermions:}
In absence of interactions, the Lindblad equation can be expressed in terms of the single-particle correlation matrix $\langle C_{pq}\rangle$ in the Heisenberg picture (see the Supplemental Information), and the equation can be solved exactly.
Figure~\ref{fig:OSIntJtotal} (a) shows the total current through the triangular lattice for different values of $t_{13}$.
The total current is not monotonic as $t_{13}$ increases.
By examining the internal currents flowing through the upper (1-2-3) and lower (1-3) branches of the triangle (Fig.~\ref{fig:illu}a), we find indeed internal current circulation in the triangle. Here the CW (CCW) circulation has an opposite current flowing along the path 1-3 (path 1-2-3).
We found CW (CCW) circulation when $t_{13}/t$ is small (large), and the bending region of the total current corresponds to the regime where all internal currents flow in the same direction.
The non-monotonic behavior of the total current as $t_{13}$ is varied can also be corroborated by the Landauer formalism~\cite{Landauer:cw,DiVentra:2008ks}. (See the SI for details.)

\begin{figure}[t]
	\begin{center}
		\includegraphics[width=1\columnwidth]{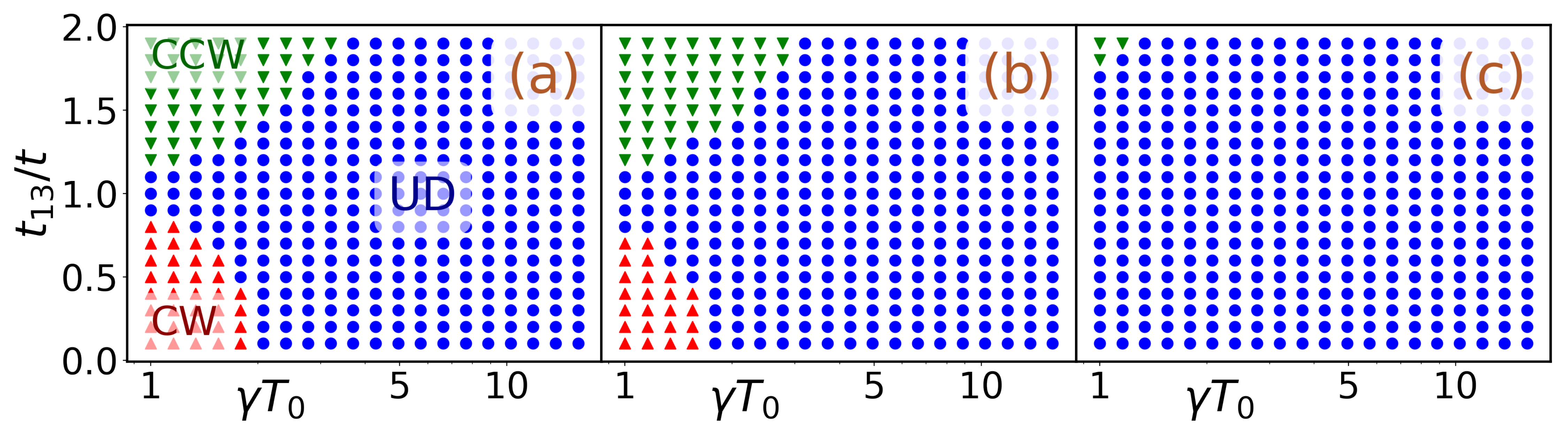}
		\caption{
			Phase diagrams of the internal current circulation from the open quantum system approach with different uniform interaction strengths $U\!=\!0$, $t$, and $5t$ from left to right.
			The blue circles indicate where all the three internal currents are unidirectional (UD), the red triangle symbols indicate where clockwise (CW) circulating current ($j_{13}\!<\!0$) can be found, and the green upside-down triangle symbols indicate where a counterclockwise (CCW) circulating current ($j_{12}\!=\!j_{23} < 0$) can be found.
		}
		\label{fig:circulation}
	\end{center}
\end{figure}

Figure~\ref{fig:circulation}(a) shows the phase diagram of the internal currents, where unidirectional, clockwise, and counterclockwise current flows are clearly distinguishable.
Spontaneous circulation of currents in quantum fluids has been found theoretically in ideal Fermi gases passing a constriction~\cite{Beria13}.
Here, we show that the circulation can be controlled in systems with a multi-connected (triangular) geometry.
By further examination, we have found that the critical point where $j_{12}$ is reversed is located at $\gamma_L\gamma_R\!=\!4(t_{13}^2-t^2)$. (See the Supplemental Information for details.)

\textit{Interacting fermions -} Having found a clear signature of internal current circulation in the TQDM, we now analyze the role of correlations.
Since the Hamiltonian, $\mathcal{H}$, consists of only three sites, we can compute numerically the dynamics of the density matrix with correlations exactly, using a fourth-order Runge-Kutta algorithm~\cite{NRE}. (See the Supplemental Information for details of the simulation.)

We first examine the system with uniform interactions $U_p\!=\!U$, $\forall p$.
The steady-state currents and their dependence on $\gamma$ are similar to the noninteracting case.
When the interaction is weak, the phase diagram showing different internal flows of currents is qualitatively the same as the diagram of noninteracting systems.
However, as the interaction becomes stronger the regimes in the parameter space showing internal current circulation shrink when compared to the noninteracting case, as shown in Fig.~\ref{fig:circulation}(b).
Therefore, by tuning the onsite interaction one can suppress internal circulation of the current.

As discussed in the noninteracting case, the total current flowing through the triangle forms a dip as $t_{13}$ varies due to a change of the current circulation.
The same behavior is found in weakly interacting systems as well.
Figure~\ref{fig:OSIntJtotal}(b) shows the total current from the left reservoir to the right one and the shaded region indicates where all internal currents are flowing in the same direction.
The total current in both the small and large $t_{13}$ regimes changes monotonically with $t_{13}$ when the internal current is circulating.
However, the total current exhibits a dip as the internal circulation changes from CCW to CW across the shaded region.
Therefore, non-monotonic behavior of the total current as $t_{13}$ varies indicates a change of the internal circulation of current in both noninteracting and weakly interacting cases.
In the strongly interacting regime, the internal circulation of current is severely suppressed as shown in Fig.~\ref{fig:circulation}(c), and the total current varies monotonically with $t_{13}$ as shown in Fig.~\ref{fig:OSIntJtotal}(c).

\begin{figure}[t]
  \begin{center}
    \includegraphics[width=0.48\textwidth]{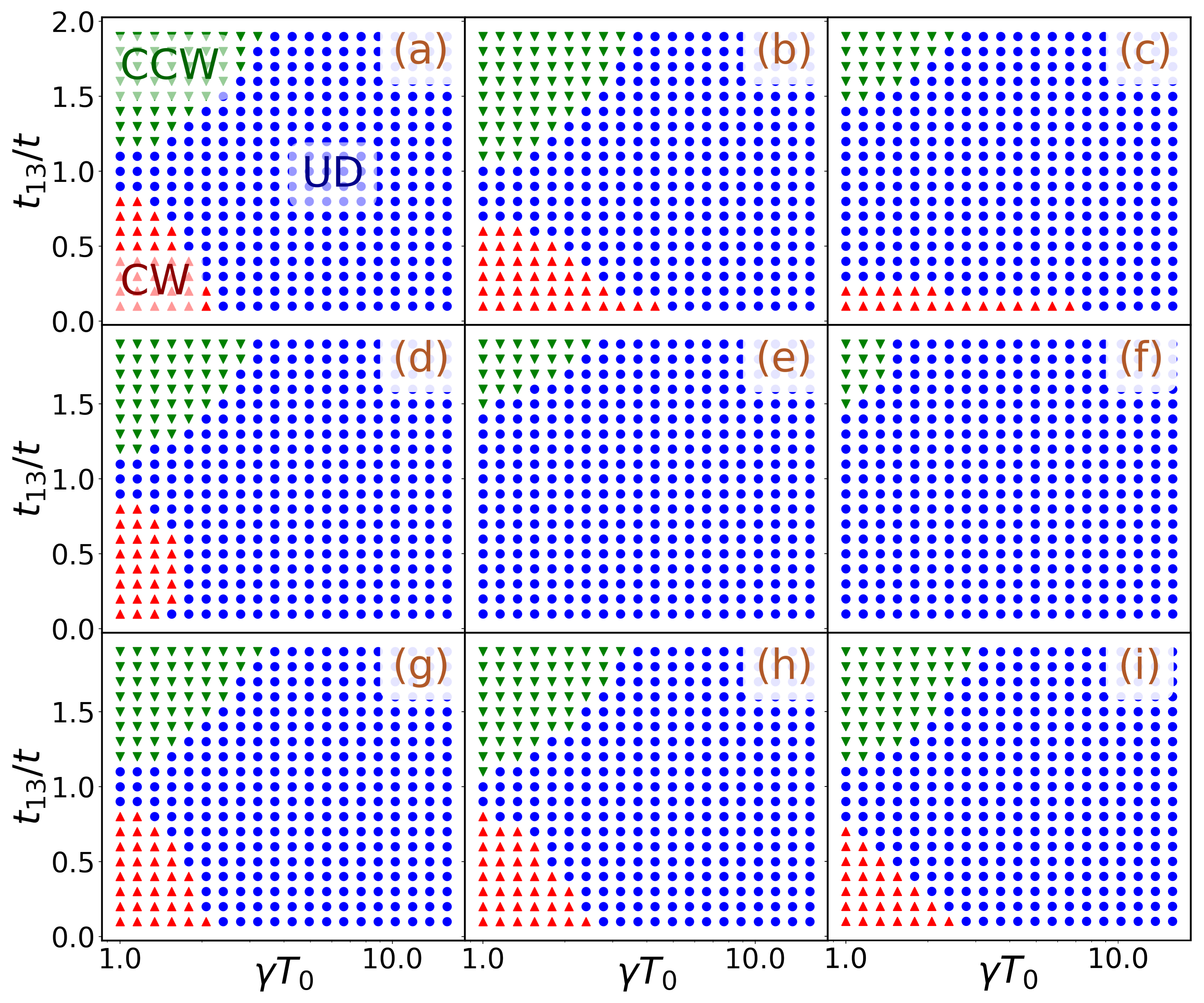}
    \caption{
      Phase diagrams of the internal current circulation with non-uniform interactions.
      The onsite repulsive interaction is only present on (a)-(c) site-$1$, (d)-(f) site-$2$, and (g)-(i) site-$3$, respectively.
			The local interaction is set to $U\!=\!0.1t$ in the left column, $U\!=\!t$ in the central column, and $U\!=\!5t$ in the right column.
			The labeling of the regimes follows the convention shown in panel (a).
    }
    \label{fig:cpU123}
  \end{center}
\end{figure}

To investigate further how correlations suppress the internal circulation of current, we assume the onsite repulsive interaction is present only on one site, while the other two sites remain noninteracting.
The phase diagrams showing where internal circulations can be found in this case are summarized in Fig.~\ref{fig:cpU123}.
The presence of interactions on site 1 affects the clockwise circulation when the interaction is strong as shown in the $U=5t$ case in Fig.~\ref{fig:cpU123}(c), but the CCW circulation is less affected.
In contrast, if the interaction is only on site 2, the suppression shown in Fig.~\ref{fig:cpU123}(e) and (f) is similar to the uniform interaction case shown in Fig.~\ref{fig:circulation}(c).
Finally, the interaction on site-3 has almost no observable influence on the circulation as shown in Fig.~\ref{fig:cpU123}(g)-(i).
Therefore, the dominant interaction effect comes from site-2, and it is possible to reduce the three-state circulation (CCW, CW, and UD) to two-state circulation (CCW and UD) as shown in Fig.~\ref{fig:cpU123}(e).

The result suggests that scattering of quantum particles is the main mechanism for tuning the internal circulation of current. That this is the case, can
be understood as follows.
In the presence of interactions on site $2$, particles are scattered from that vertex, so the current flowing through the upper (1-2-3) path is reduced.
This makes the CW circulation unfavorable because it requires a large current through the upper path and a counter-flowing current on the lower (1-3) path.
On the other hand, adding scattering mechanisms like onsite interaction to site $1$ or $3$ leaves the phase diagrams intact (or completely suppresses the internal circulations).
Similar results occur if one includes onsite attractive potentials (see the SI), thus confirming the single-particle scattering mechanism we have just described.
Moreover, introducing inhomogeneous hopping coefficients, for example by setting $t_{12}\neq t_{23}$, leads to additional scattering along the upper path and also shifts the boundary between different types of circulation on the phase diagrams.
Inhomogeneous interactions or onsite potentials may be achievable in quantum dots coupled to cavities by tuning the photon-exciton coupling.
In this respect, the photonic-circuit structure shown in Fig.~\ref{fig:illu}(c) has an advantage over the electrostatic quantum dots when it comes to configurations with tunable inhomogeneity.

\textit{Conclusions -} We have considered a triangular quantum-dot metastructure connected to two reservoirs and studied the relation between its steady-state total current and its internal current circulation.
Internal circulation of current in the triangle are discovered in both closed- and open- system approaches, and the direction can be tuned by a variety of parameters including the hopping coefficient, local interactions or potential, and system-reservoir coupling. Therefore, the internal current circulation is model-independent. Moreover,
the overall current exhibits non-monotonic behavior when the the circulation reverses.
The phase diagrams showing how the circulation can be tuned will assist designs of quantum devices utilizing the internal circulation of current.

Importantly, our findings are not limited to quantum dots because the generic formalism based on quantum mechanics establishes the robustness of the internal current circulation in quantum transport.
It is also possible to use the recently developed lattice fermion simulators based on superconducting elements~\cite{Barends:jc} or ultracold atoms in engineered reservoirs and constrictions~\cite{Lebrat18} to explore similar transport phenomena in other controllable quantum systems.

\textit{Acknowledgment:} We thank Mekena Metcalf for useful discussions.
C.Y.L. acknowledges the support from the U.S. Department of Energy through the Center for Integrated Nanotechnologies, a basic energy science user facility.
M.S. acknowledges support from the Defense Threat Reduction Agency (Grant No. HDTRA1-15-1-0011) and the Air Force Office of Scientific Research (Grant No. FA9550-16-1-0278).
The simulations were performed by the MERCED Cluster in UC Merced supported by the National Science Foundation (Grant No. ACI-1429783).


\pagebreak

\begin{widetext}
\begin{center}
	\textbf{Supplemental Information: Tunable current circulation in triangular quantum-dot metastructures}
\end{center}
\end{widetext}
\beginsupplement

\section{Experimental details of triangular triple quantum dot}
For the cavity-quantum dot hybrid system shown in Fig. 1(c) in the main text, methods exist to locally tune each cavity, making mode-matching between all three cavities feasible. For example, Faraon {\it et al.} report that photo-darkening of a thin chalcogenide glass layer deposited on top of the device can be used to locally tune cavity modes by about 3nm at 940nm ~\cite{Faraon08}.
In order to realize and control the photon blockade regime the QDs’ excitonic transitions must also be fine-tuned into resonance with the modes of the photonic crystal cavities.
Such tuning can be achieved independently from the cavity tuning via the quantum confined Stark effect, by embedding the dots in p-i-n-type electric field-effect structures.
Here, the use of stacked quantum dot pairs is advantageous, as either the dots in one layer can be tuned separately from those in the other layer~\cite{Kim09}, or interdot transitions of coupled QD (CQD) can be used.
With CQDs the electrical tuning of the interdot QD transitions is much enhanced~\cite{Scheibner09,Kerfoot14}. Tuning ranges of tens of meV~\cite{Scheibner09,Kerfoot14} allow for low QD density (1 QD/$\mu m^2$ or 1 QD per cavity) to be used.

\begin{figure}[t]
	\begin{center}
		\includegraphics[width=0.2\textwidth]{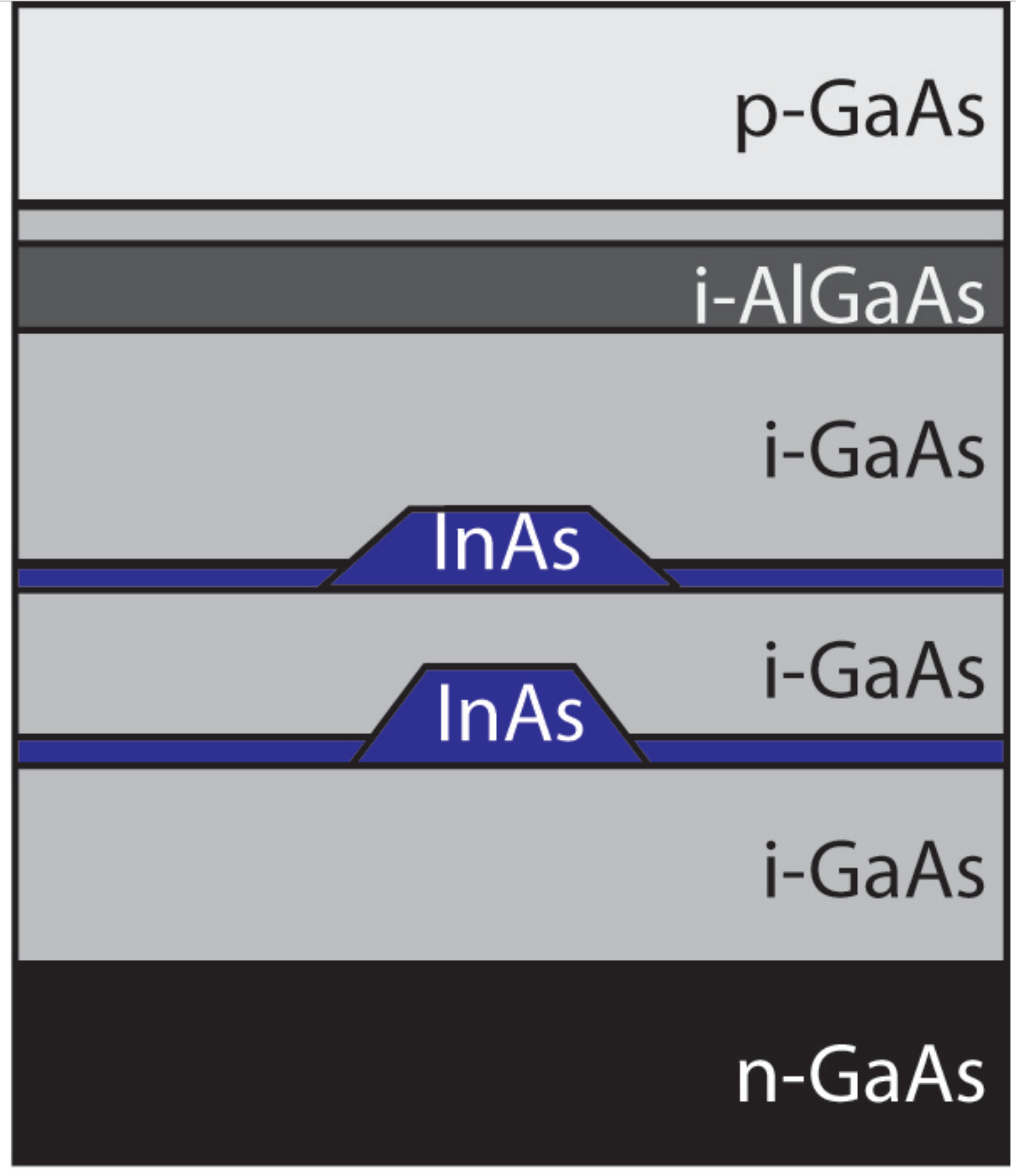}
		\caption{
			Schematic layer sequence of the underlying pin-diode structure for Fig. 1(c) in the main text.
		}
		\label{fig:layer}
	\end{center}
\end{figure}

Individual electrical tuning of the excitonic transitions of each of the three dot-cavity systems may be achieved by etching through the top doped layer of the p-i-n-type structure along lines of holes (see Figure 1(c) in the main text and Figure~\ref{fig:layer}), allowing separate potentials, $V_1$, $V_2$, $V_3$, to be applied.
Tuning the excitonic transitions helps accomplish two things. First, it allows to establish strong coupling between the QD exciton and the cavity mode, resulting in a spectral shift of the polariton state and hence introducing non-linearity in the photonic response of the cavity. Second, how well the three cavity-dot systems are matched spectrally determines the coupling strength between them. The coupling between the three dot-cavity systems may additionally be controlled in other ways. For example, the aforementioned local photo-darkening method may be applied in the region between the cavities. Likewise, the photonic crystal may be modified by adsorption of chemicals~\cite{Strauf06}, or strain can be used to shift the excitonic transitions~\cite{Carter17}. Conversely, the three dot-cavity systems may be used for sensing adsorption of chemicals or strain generating forces.

\section{Landauer approach of noninteracting fermions}
The steady-state current of noninteracting fermions going through a junction can be obtained from the Landauer approach~\cite{Landauer:cw,DiVentra:2008ks}.
Here we apply this method to a three-site triangular lattice connected to two leads modeled as a closed or open system as illustrated in Fig.~\ref{fig:illu}.
The Hamiltonian of the triangular lattice is
\begin{equation}\label{eq:Htri}
\mathcal{H}_{\text{tri}}=-t(c^\dagger_1c_2+c^\dagger_2c_3+h.c.)-t_{13}(c^\dagger_1c_3+c^\dagger_3c_1),
\end{equation}
where $c^\dagger_n$ ($c_n$) is the fermion creation (annihilation) operator on site-$n$, and we assume $t_{12}=t_{23}=t$. Relaxing the latter condition only leads to quantitative changes of the results.
We will first present the result from Landauer formula and then use numerical simulations to analyze the details.

\subsection{Landauer formula}
Following the standard procedures~\cite{DiVentra:2008ks}, the steady-state current of one spin species from the Landauer formula is given by
\begin{equation}\label{eq:landauer}
J=\frac{e}{2\pi \hbar}\int dE [f_L(E)-f_R(E)]T(E).
\end{equation}
Here $f_{L,R}(E)$ is the particle distribution of the left or right reservoir, and $T(E)$ is the transmission coefficient of the junction.
To apply the formula to the system shown in Fig.~\ref{fig:illu}(b), we model the two reservoirs as two semi-infinite uniform lattices with hopping coefficient $t^\prime$. We will choose $e\equiv 1$ and $\hbar=1$.

The transmission coefficient in Eq.~\eqref{eq:landauer}, $T(E)$, can be obtained from $Tr[\Gamma_R G_3^+ \Gamma_L G_3^-]$.
Here the Green's function of the coupled system is $G_3^+(E)=[E-H_3-\Sigma_L -\Sigma_R]^{-1}$ and $G_3^- =(G_3^+)^*$, where
$$
H_3=\left(\begin{array}{ccc}
0 & -t & -t_{13} \\
-t & 0 & -t\\
-t_{13} & -t & 0
\end{array} \right)
$$
is the Hamiltonian of the triangle.
The Green's functions of the reservoirs are $G_{L,R}(E)=[E-(E-i\sqrt{4t^2-E^2})/2]^{-1}$, and the two reservoirs are connected to the system through hopping coefficient $t^\prime$ which have the self-energy
$$\Sigma_{L}=\left(\begin{array}{ccc}
G_{L}t^{\prime 2} & 0 & 0 \\ 0 & 0 & 0\\ 0 & 0 & 0
\end{array} \right)\text{ , and }
\Sigma_{R}=\left(\begin{array}{ccc}
0 & 0 & 0 \\ 0 & 0 & 0\\ 0 & 0 & G_{R}t^{\prime 2}
\end{array} \right).
$$
Finally, $\Gamma_{L}=\left(\begin{array}{ccc} A_L & 0 & 0 \\ 0 & 0 & 0\\ 0 & 0 & 0 \end{array} \right)$ and $\Gamma_{R}=\left(\begin{array}{ccc} 0 & 0 & 0 \\ 0 & 0 & 0\\ 0 & 0 & A_R \end{array} \right)$ with $A_{L,R}=i(G_{L,R}-G^*_{L,R})=\sqrt{4t^2-E^2}/t^2$ corresponding to the coupling of the triangular lattice to the two reservoirs.


\begin{figure}[t]
	\begin{center}
		\includegraphics[width=0.47\textwidth]{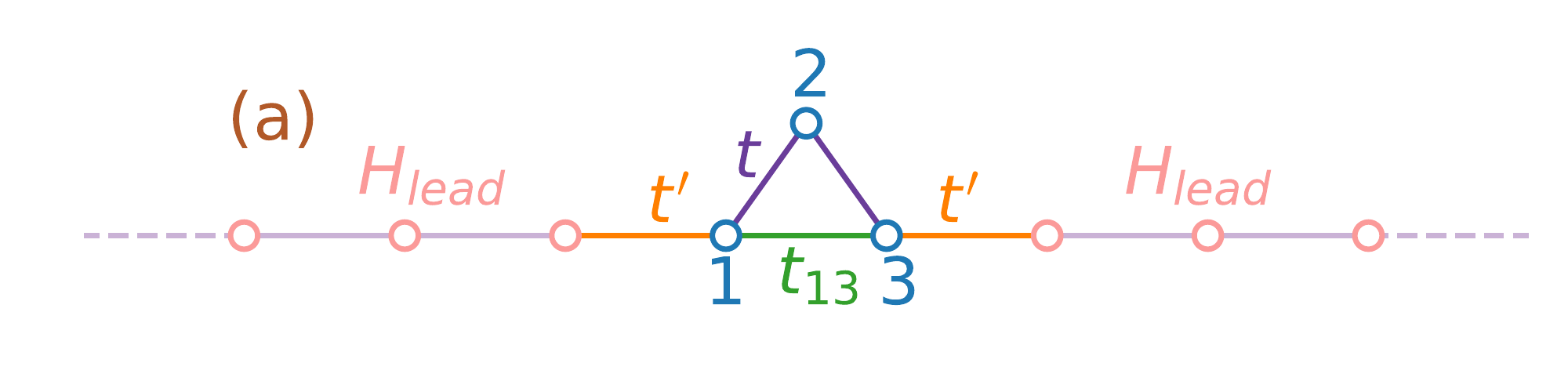}
		\includegraphics[width=0.47\textwidth]{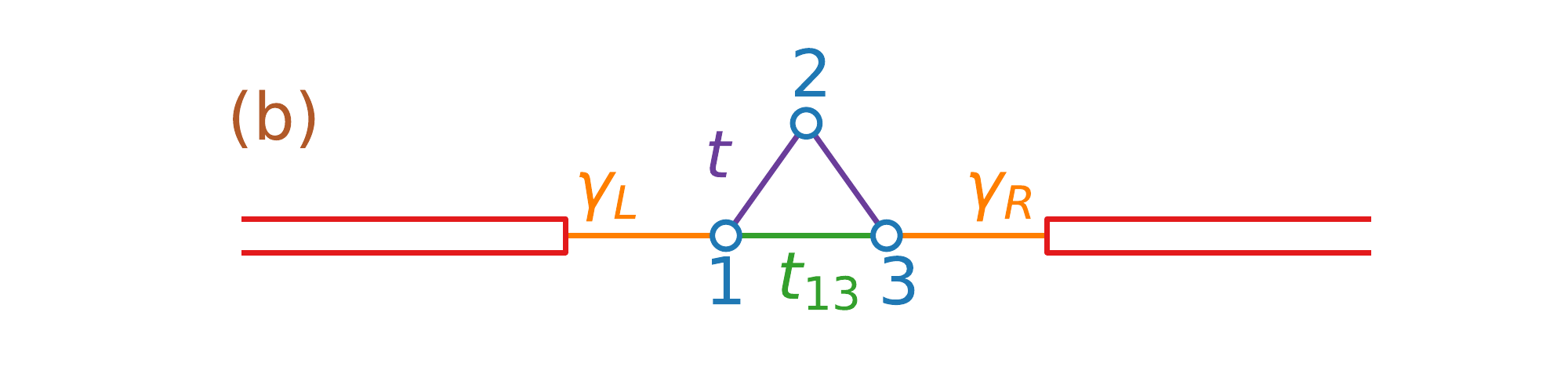}
		\caption{
			Illustration of a triangular triple quantum dot connected to two different kinds of particle reservoirs.
			(a) Microcanonical picture of transport: An isolated quantum system with two finite-size lattices serving as reservoirs.
			(b) An open quantum system approach with external reservoirs.
		}
		\label{fig:illu}
	\end{center}
\end{figure}

\begin{figure}[t]
	\begin{center}
		\includegraphics[width=0.45\textwidth]{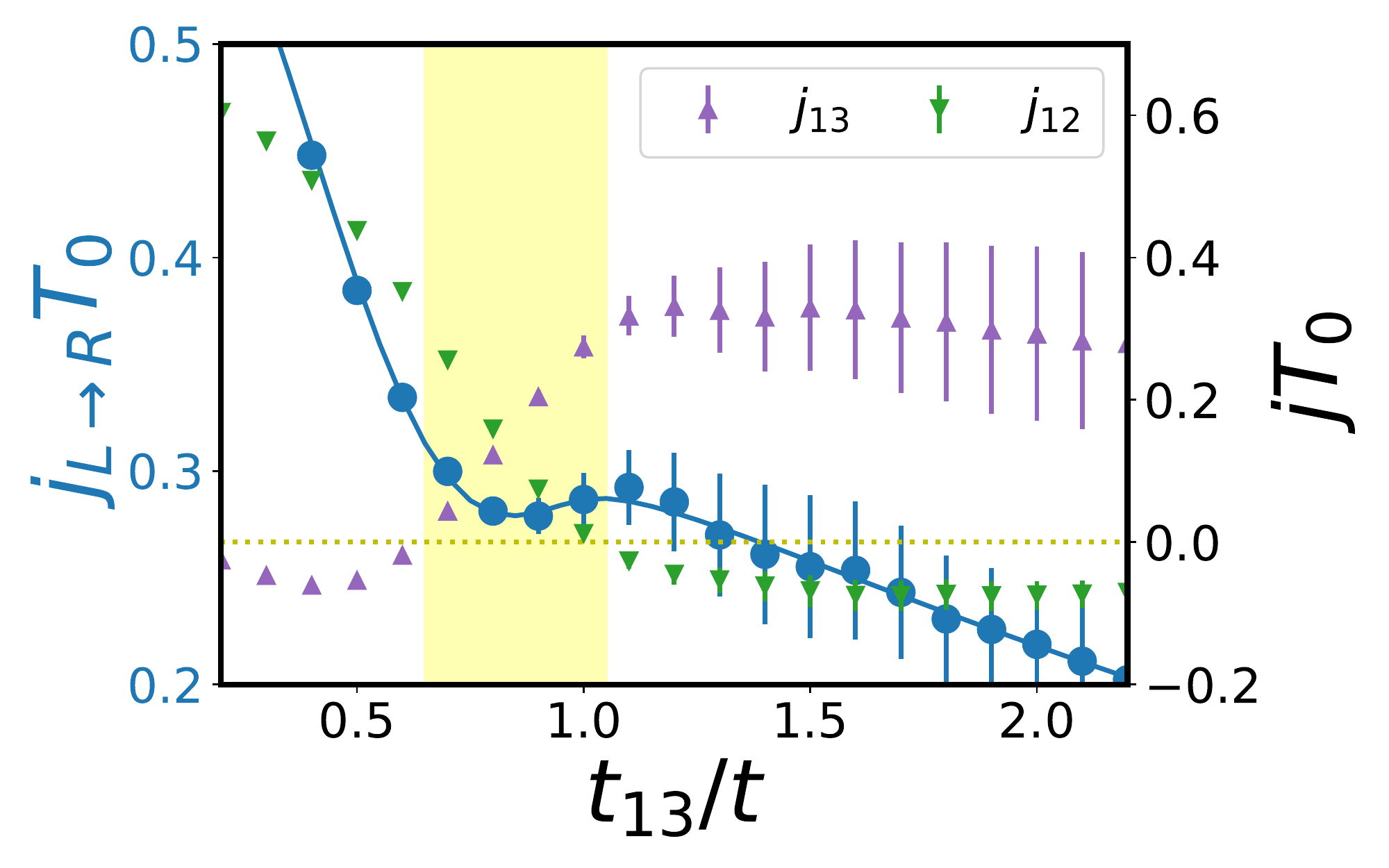}
		\caption{
			The quasi-steady state currents from the microcanonical picture of transport versus different $t_{13}$ with $t^\prime=t$.
			The total current from the left to the right, $j_{L\rightarrow R}$ (blue circles), is defined on the link $t^\prime$ connecting the system and one of the reservoirs.
			The shaded regime indicates the currents on the triangle ($j_{12}$, $j_{13}$) are uni-directional (all from the left to the right), and the total current exhibits a non-monotonic behavior when the internal circulation on the triangle changes direction.
			In the left (right) of the shaded region the circulation is clockwise (counterclockwise).
			The error bars are determined from the time average of the quasi-steady state values. The solid line shows the result from Landauer formula~(\ref{eq:landauer}) at zero temperature.
		}
		\label{fig:ndc}
	\end{center}
\end{figure}

We will focus on quantum transport through the triangular lattice and consider only the zero-temperature limit.
As the temperature approaches $0$, the Fermi distributions $f_{L,R}$ become step functions, and the current is $J=\int_{-2t}^{2t}dE T(E)/(2\pi)$ if the left reservoir is fully occupied and the right one is empty.
Figure~\ref{fig:ndc} shows the current from Landauer formula with $t^{\prime}=t$.
Even though the link between site 1 and site 3 in Fig.~\ref{fig:illu} looks like a shortcut, its presence actually suppresses the current rather than enhancing it as one can see in Figure~\ref{fig:ndc}.
This is because additional paths may introduce scattering of the wavefunctions of quantum particles at the junction and reduce the overall tunneling probability through the junction.
Another feature shown in Fig.~\ref{fig:ndc} is that the current exhibits non-monotonic behavior as $t_{13}$ increases.
The origin of such non-monotonic behavior will be revealed in the next section.

\subsection{Microcanonical picture of transport}
To analyze the details of the transport through the triangular lattice, we simulate the system shown in Fig.~\ref{fig:illu}(a) by modeling the two reservoirs as long but finite lattices with uniform tunneling coefficients, namely we adopt a microcanonical picture of transport~\cite{DiVentra:2004bx}.
Initially, the left reservoir is filled up completely while the right one is completely empty.
The two sides start exchanging particles at $T\!=\!0$.
The Hamiltonians of the reservoirs are of the one-dimensional free fermion form:
\begin{eqnarray}
\mathcal{H}_{\alpha,\text{lead}}=-\sum_{\langle ij\rangle}t(a^\dagger_{\alpha,i}a_{\alpha,j}+h.c.),\nonumber
\end{eqnarray}
where $\langle ij\rangle$ denotes nearest neighbors and $\alpha\in\text{L}$ or R for  the left or right reservoir, respectively. The time unit is $T_0=\hbar/t$.
The links connecting the reservoirs and the triangle are set to be the same on both sides,
\begin{eqnarray}
\mathcal{H}_{\text{couple}}=-t^\prime(a^\dagger_{L,N}c_{1}+a^\dagger_{R,1}c_{3}+h.c.).
\end{eqnarray}

\begin{figure}[t]
	\begin{center}
		\includegraphics[width=0.4\textwidth]{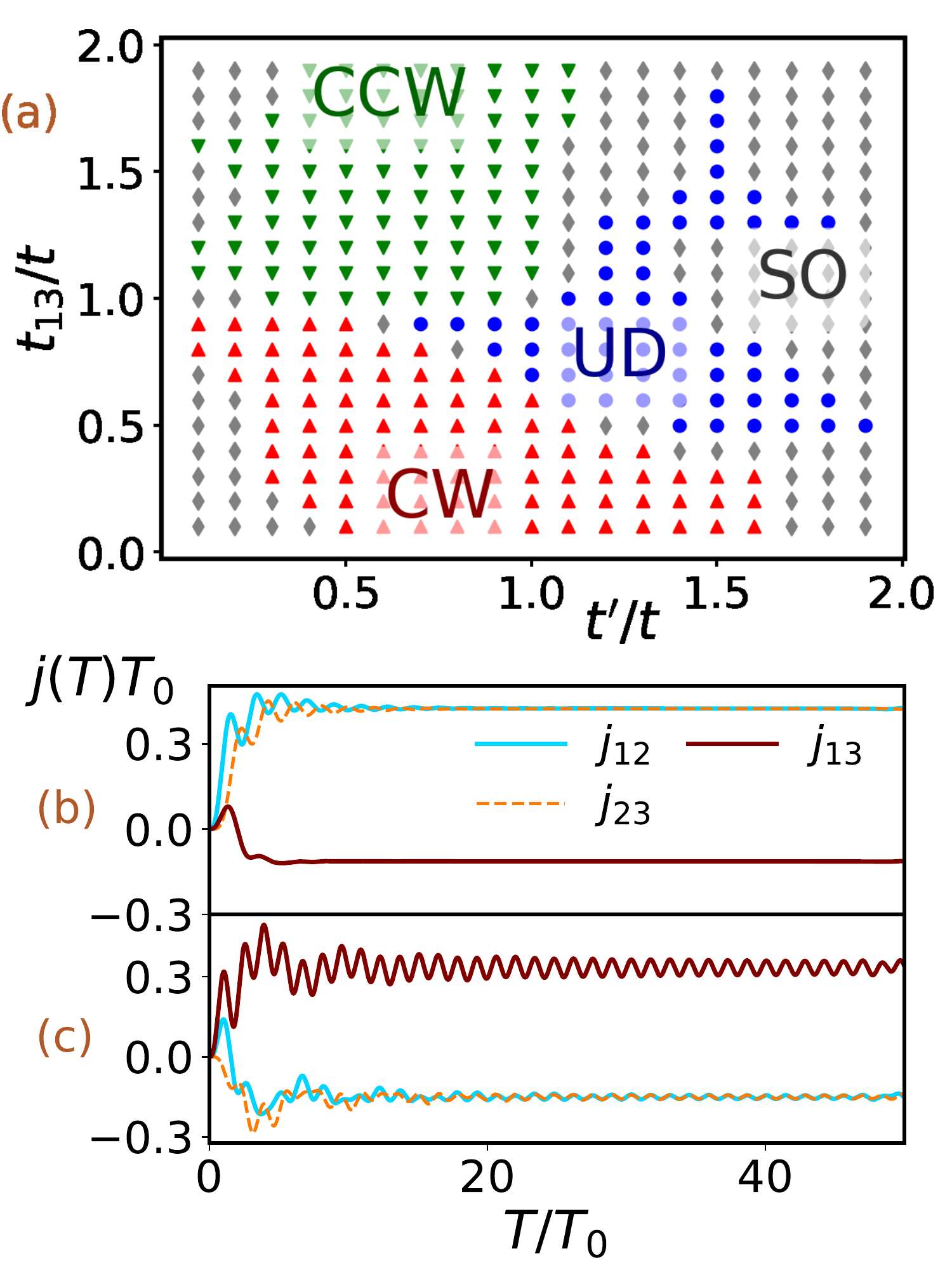}
		\caption{
			(a) Phase diagram showing different internal current circulations of an isolated system. The blue circles indicate where all three currents are unidirectional (UD), the red up-triangles indicate where clockwise (CW) circulating current ($j_{13}\!<\!0$) can be found, and the green down-triangles indicate where counterclockwise (CCW) circulating current ($j_{12}\!=\!j_{23}\!<\!0$) can be found. If the currents exhibit strong oscillations (SO), the regime is marked by gray diamonds.
			Details of the current versus time of a system with clockwise (counterclockwise) circulation and $t_{13}\!=\!0.5t$ ($t_{13}\!=\!1.5t$) are shown in (b) [(c)].
			Here $t^\prime\!=\!0.7t$ in both cases.
		}
		\label{fig:CS}
	\end{center}
\end{figure}

The particles start to flow to the right reservoir through the triangle, and a quasi-steady state current (QSSC) on each link of the triangle can be observed.
The QSSC corresponds to a plateau when the current is plotted versus time, and Fig.~\ref{fig:CS}(d) provides some examples.
The duration of the QSSC scales linearly with the reservoir size $L$.
In our simulations we use $L\!=\!50$ lattices sites for each reservoir and the value of the QSSC is insensitive to $L$ when the latter is sufficiently large.
The dynamics are simulated numerically by computing the single-particle correlation matrix $\langle C_{pq}\rangle=\langle c^{\dagger}_p c_q\rangle$ with the equation of motion
\begin{equation}
\frac{d}{dT}\langle C_{pq} \rangle =-i\langle[C_{pq},\mathcal{H}_{total}]\rangle,\nonumber
\end{equation}
where $\mathcal{H}_{total}\!=\!\sum_\alpha\mathcal{H}_{\alpha,\text{lead}}+\mathcal{H}_{\text{tri}}+\mathcal{H}_{\text{couple}}$ is total Hamiltonian of the entire isolated system.
The quasi-steady state current on the link from site $p$ to site $q$ of the triangular lattice can be extracted from the off-diagonal elements of the single-particle correlated matrix as
\begin{equation}
j_{pq}=-2t_{pq}\text{Im}\langle C_{pq}\rangle.
\end{equation}
In a closed system, the current on each link still oscillates even in the long time limit due to finite-size effect. Moreover, the current will reverse its direction when the wavefunction hits the boundary because the system is finite. Therefore, the results are averaged over a time period when the system is in a quasi-steady state before the revival. Explicitly,
\begin{equation}
\overline{j_{pg}}=\sum_{T_s=T_i}^{T_f}j_{pq}(T_s)\frac{\delta T}{T_f-T_i},
\end{equation}
where $\delta T\!=\!0.005T_0$ is the time step used in the Runge-Kutta simulations~\cite{NRE}.
The results are taken between $T_i\!=\!20T_0$ and $T_f\!=\!30T_0$ where the current signal is oscillating around a stable average.

By varying $t_{13}$ within the triangle and the coupling $t^\prime$ between the system and leads, the system results in different QSSC behavior.
One particular example is shown in Fig.~\ref{fig:ndc}.
The overall current flowing from the left reservoir to the right can reach steady states but the currents on the internal links of the triangle may not exhibit steady states.
In certain parameter range, the current exhibits strong oscillations which cause a large statistical error as an error bar shown in Fig.~\ref{fig:ndc}.
For those regimes where steady state currents in the links of the triangular lattice can not be properly identified, we mark it as gray diamonds in Fig.~\ref{fig:CS}(a).
Later on we will show the open quantum system approach always leads to steady-state current in the internal links due to strong decoherence through the different modeling of the coupling to the environment.
Importantly, we identify internally circulating currents within the triangular lattice in certain parameter range.
The circulation can be clockwise or counterclockwise depending on the parameters.
Figure~\ref{fig:CS}(a) shows a phase diagram indicating where clockwise, counterclockwise, and unidirectional currents can be found.
Two examples showing circulating currents are presented in Figs.~\ref{fig:CS}(b) and (c).

\section{Quantum Master Equation Approach}\label{sec:os}
We can rewrite the Lindblad equation, Eq. (3) in the main text, in the Heisenberg picture for an operator $\mathcal{O}$ as
\begin{eqnarray}
\frac{d}{dT}\langle\mathcal{O}\rangle&=&-i\langle\left[\mathcal{O},\mathcal{H}_{\text{tri}}\right]\rangle
+\gamma_L\langle\left(c_1\mathcal{O}c^\dagger_1-\frac{1}{2}\{c_1c^\dagger_1,\mathcal{O}\}\right)\rangle \nonumber \\
& &	+\gamma_R\langle\left(c^\dagger_3\mathcal{O} c_3-\frac{1}{2}\{c^\dagger_3c_3,\mathcal{O}\}\right)\rangle,
\end{eqnarray}
where the Hamiltonian is given by Eq. (2) of the main text.

\subsection{Noninteracting fermions}
For noninteracting fermions, we take the single-particle correlation matrix $\langle C_{pq}\rangle\!=\!\langle c^\dagger_pc_q\rangle$ and simulate the dynamics by reducing the Lindblad equation to the following form:
\begin{eqnarray}\label{eq:lindblad}
\frac{d}{dT}\langle C_{pq} \rangle&=&-i\langle[C_{pq},\mathcal{H}_{\text{tri}}]\rangle
+ \gamma_L\langle c_1c^\dagger_pc_qc^\dagger_1-\frac{1}{2}\{c_1c^\dagger_1,c^\dagger_pc_q\}\rangle \nonumber \\
& &	+ \gamma_R\langle c^\dagger_3c^\dagger_pc_qc_3-\frac{1}{2}\{c^\dagger_3c_3,c^\dagger_pc_q\}\rangle,
\end{eqnarray}
where the first term can be expressed as a combination of $\langle C_{pq}\rangle$.
By using the Wick decomposition, all the terms can be expressed in terms of the elements of $\langle C_{pq}\rangle$.
Thus, the equation becomes a set of coupled differential equations of nine distinguishable elements.
Looking for a steady state, namely by setting $d\rho/dT\!=\!0$, this set of equations can be solved exactly.

The diagonal elements $\langle C_{pp}\rangle$ are real numbers and represent the density on the three sites of the triangular lattice, and the current on each link can be obtained from the imaginary part of the off-diagonal elements. In the steady-state,
\begin{eqnarray}\label{eq:OSSolution}
j_{12} &=& j_{23} = 2t\text{Im}\langle C_{12} \rangle\nonumber\\
&=&\frac{4t^2\gamma_L\gamma_R[t\gamma_L\gamma_R-4t(t_{13}^2-t^2)]}{\mathcal{D}(\gamma_L+\gamma_R)}, \\
j_{13} &=& 2t_{13}\text{Im}\langle C_{13}\rangle\nonumber\\
&=&\frac{4tt_{13}\gamma_L\gamma_R[t_{13}\gamma_L\gamma_R+4t_{13}(t_{13}^2-t^2)]}{\mathcal{D}(\gamma_L+\gamma_R)},
\end{eqnarray}
where $\mathcal{D}=t\gamma_L^2\gamma_R^2+8t(t_{13}^2+t^2)\gamma_L\gamma_R+16t(t_{13}^2-t^2)^2$.
Therefore, the locations where the internal currents reverse can be identified. We remark that setting $t_{12} \neq t_{23}$ only modifies the analysis quantitatively. Therefore, introducing inhomogeneous hopping coefficients only shifts the boundaries on the phase diagrams.

\subsection{Interacting fermions}
We model interacting fermions as the Hubbard model shown in Eq. (2) in the main text.
In this work we only consider a single-band model with the Coulomb interaction approximated by an onsite repulsive interaction.
Therefore, the triangular triple quantum dot is described by a fermion Hubbard model and the exchange of fermions with the reservoirs is described by the Lindblad equation.

Unlike the noninteracting case, the equation of motion of $\langle C_{pq}\rangle$ cannot be reduced to a closed set of equations.
Instead, we cast the density matrix $\rho$ in the Fock-space basis and use the Lindblad equation to monitor its time evolution.
In the Fock-space basis $\vert\xi\rangle \!=\! \vert s_1, s_2, s_3\rangle$ counting the occupation number $s_n$ on site $n$, there are four states $\vert s_i\rangle\in\{\vert0\rangle, \vert\downarrow\rangle, \vert\uparrow\rangle, \vert\uparrow\downarrow\rangle\}$ on each site.
The density matrix in the Fock-space basis is
$\rho=\sum_{\xi,\xi^\prime}\varrho_{\xi^\prime\xi}\vert \xi^\prime\rangle \langle \xi\vert$.
Here, $\varrho_{\xi^\prime\xi}$ is a $4^3$ by $4^3$ matrix. In the same basis, we can represent the TQDM Hamiltonian, Eq. (2) in the main text, as a $4^3$ by $4^3$ matrix.
In this representation, both the hopping terms and the onsite interactions are treated {\it exactly}. For example,
\begin{eqnarray}
\rho Un_{\uparrow,1}n_{\downarrow,1}&=&
\sum_{\xi,\xi^\prime}U\varrho_{\xi^\prime\xi} \vert \xi^\prime\rangle \langle \xi\vert n_{\uparrow,1}n_{\downarrow,1},
\end{eqnarray}
where the only non-zero contribution comes from $\vert \xi^\prime (\xi)\rangle\!=\!\vert\uparrow\downarrow, s_2, s_3\rangle$.
Similar operations can be carried out for the hopping terms as well as the onsite potentials.
Then, solving the Lindblad equation with the Runge-Kutta method~\cite{NRE} leads to the steady-state density matrix.
In our simulations of the open quantum system, the density matrix always evolves into a steady state. Moreover, the steady state is insensitive to different initial conditions.
The expectation value of an operator $\hat{O}$ is given by $\langle \hat{O} \rangle \!=\! \text{Tr}(\rho\hat{O})$, where $\text{Tr}$ denotes the trace.
Taking the operator $\sum_{\sigma}c^\dagger_{1\sigma}c_{2\sigma}$ as an example, one obtains
$\sum_\sigma\text{Tr}(\rho c^\dagger_{1\sigma}c_{2\sigma})
= \sum_\sigma\sum_{\xi,\xi^\prime} \varrho_{\xi^\prime\xi} \langle \xi \vert c^\dagger_{1\sigma}c_{2\sigma} \vert \xi^\prime \rangle$.
The current $j_{pq}$ from site $p$ to site $q$ can be evaluated by using Eq. (4) in the main text once the steady-state $\rho$ is found from the Lindblad equation.

\subsection{Noninteracting fermions with attractive potential}
\begin{figure}[t]
	\begin{center}
		\includegraphics[width=0.48\textwidth]{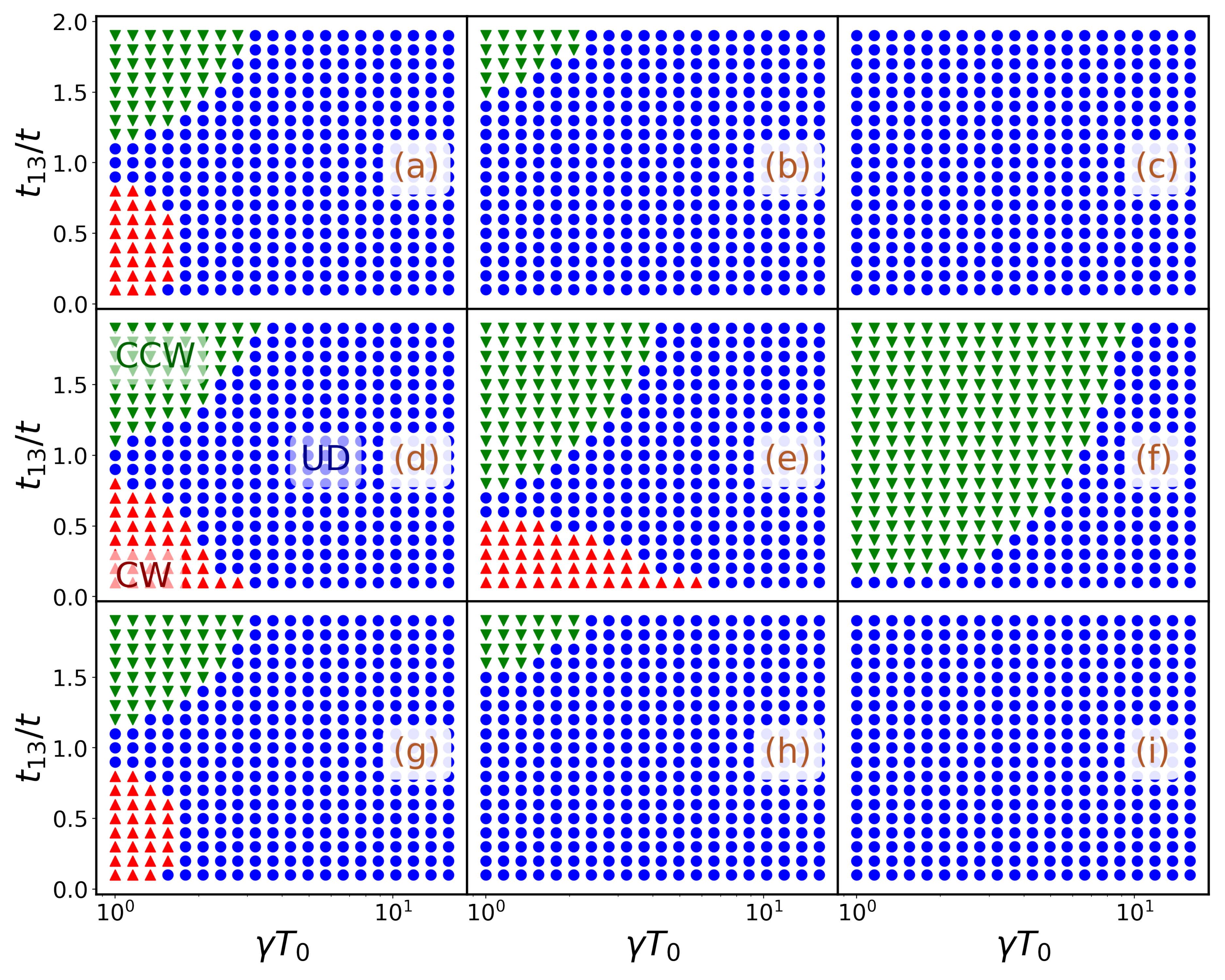}
		\caption{
			Phase diagrams of the internal current circulation of noninteracting fermions in the presence of an attractive potential only on site-1 [(a) - (c)], only on site-2 [(d) - (f)], and only on site-3 [(g) - (i)].
			The magnitude of the potential, $\tilde{V}$, is $-0.1t$ in the left column, $-t$ in the middle column, and $-10t$ in the right column. A quantum master
			equation approach has been used to generate this plot.
		}
		\label{fig:OS2V}
	\end{center}
\end{figure}

To explore other possibilities of tuning the internally circulating current and to elucidate its physical origin, we return to noninteracting fermions with an attractive onsite potential on a selected site.
Such an onsite potential may be induced by a gate voltage localized to only one quantum dot.
While the onsite repulsive interactions considered previously increase the energy, the attractive potential does the opposite and it can help clarify how energy shifts affect transport.
We introduce the Hamiltonian
\begin{equation}
\mathcal{H}=\sum_\sigma\mathcal{H}_{\text{tri},\sigma}-\tilde{V}(n_{j,\uparrow}+n_{j,\downarrow}).
\end{equation}
Here the onsite potential $\tilde{V}$ only applies to site $j$.
The phase diagrams of internal circulations are shown in Fig.~\ref{fig:OS2V} for different sites with selected values of $\tilde{V}$.

The attractive onsite potential clearly suppresses the CW circulation as shown in Fig.~\ref{fig:OS2V}.
However, the fate of the CCW circulation depends on the location of the onsite potential.
If the potential is on site-1 or site-3, the internal circulation will be completely suppressed as the potential increases, as shown in the top and bottom rows of Fig.~\ref{fig:OS2V}.
In contrast, if the potential is on site-2, the CCW circulation will survive even when $U/t$ is large, and a large portion of the parameter space is occupied by the CCW circulation.
This implies that one can tune the three-state circulation (CW, CCW, and unidirectional) to two-state circulation (CCW and unidirectional) by introducing a potential on site-2, and the situation is similar to the case with only repulsive interactions on site-2 as discussed in the main text.


\end{document}